\documentclass[twocolumn,noshowpacs,preprintnumbers,amsmath,amssymb]{revtex4-1}

\usepackage{graphicx}
\usepackage{dcolumn}
\usepackage{bm}

\newcommand\vex[1]{\mathbf{#1}}

\def\dd{\mathrm{d}}

\def\Kp{\vex K_+}
\def\Km{\vex K_-}
\def\Kpm{\vex K_\pm}

\def\highlight#1{{#1}}

\begin{document}

\title{Floquet-Engineered Valleytronics in Dirac Systems}

\author{Arijit Kundu}
\affiliation{Department of Physics, Indiana University, Bloomington, IN 47405}
\author{H.A. Fertig}
\affiliation{Department of Physics, Indiana University, Bloomington, IN 47405}
\author{Babak Seradjeh}
\affiliation{Department of Physics, Indiana University, Bloomington, IN 47405}

\begin{abstract}
Valley degrees of freedom offer a potential resource for quantum information processing
if they can be effectively controlled.  We discuss an optical approach to this
problem in which intense light breaks electronic symmetries of a
two-dimensional Dirac material.
The resulting quasienergy structures may then differ for different
valleys, so that the Floquet physics of the system can be exploited to
produce highly polarized valley currents. This physics can be utilized to realize a valley valve whose behavior is determined optically. We propose a concrete way to
achieve such valleytronics in graphene as well as in a simple model
of an inversion-symmetry broken Dirac material. We study the effect numerically and demonstrate its robustness
against moderate disorder and small deviations in optical parameters.
\end{abstract}

%\date{\today}
\maketitle

\emph{Introduction.}---%
Since the advent of graphene as a two-dimensional electronic material which can be
produced in the laboratory~\cite{Novoselov_2004,Castro_Neto_RMP},  the possibility of exploiting the valley degree of freedom within it~\cite{rycerz_2007b} and other Dirac systems has been vigorously studied. An important component of such valleytronic systems is the transport and detection of valley currents.
Many of the ideas proposed to do so involve controlling the structure of the system,
either using specific edge structures
~\cite{rycerz_2007b} or bulk nanostructures~\cite{song_2013,grujic_2014}.
These ideas are limited by the precision they require to control the nanostructure.
A particularly interesting way around these limitations combines intrinsic band properties with optics to yield
valley-contrasting behavior.
In gapped Dirac systems such as MoS$_2$ and WS$_2$, or bilayer graphene
in a perpendicular electric field, valley currents can be induced using the differing Berry's curvatures of
the valleys~\cite{niu_2008,abergel_2009,mak_2014,gorbachev_2014,qi2014,SieMcILee15a}.
In such systems, circularly polarized light can excite different electron-hole pair populations in different
valleys~\cite{cao_2012,zeng_2012,shan_2015}., leading to a ``valley Hall effect''
~\cite{mak_2014,gorbachev_2014,lensky_2014,sui_2015,shimazaki_2015}.

%----- Fig. 1 -----%
\begin{figure}[t]
\includegraphics[width=3.2in]{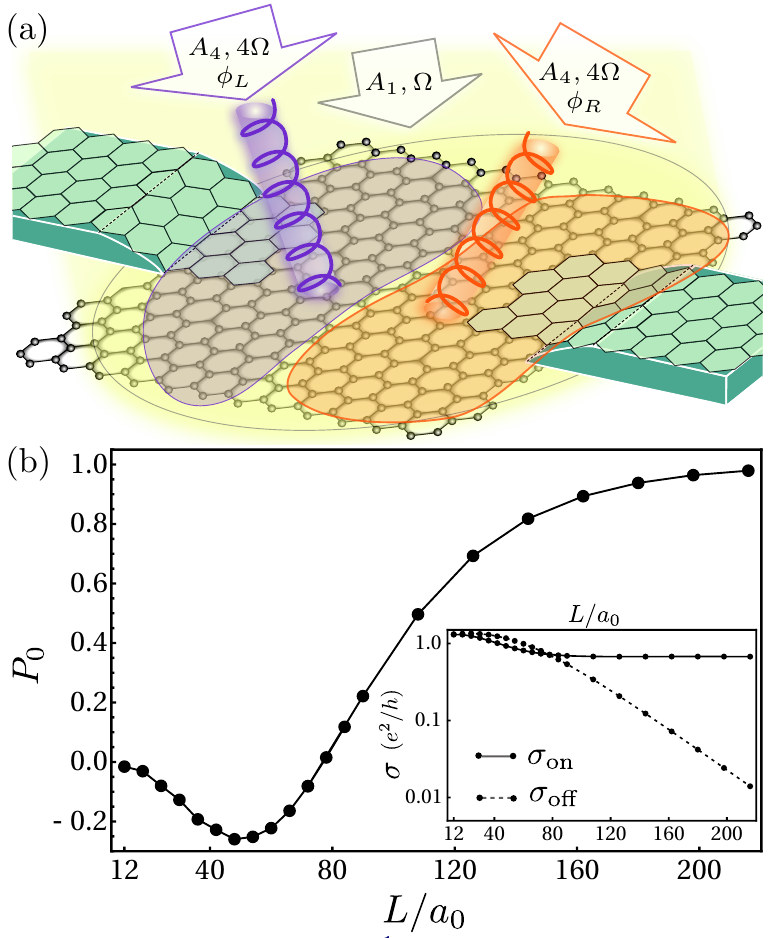}
\caption{ (color online)
(a) The schematics of the graphene system: A background laser with frequency $\Omega$ and intensity $(c/4\pi)A_1^2\Omega^2$ is supplemented with a circularly-polarized 4th harmonic of frequency $4\Omega$ and intensity $(4c/\pi)A_4^2\Omega^2$ with a relative phase offset $\phi_L$ on the left and $\phi_R$ right half of the device. (b) The valley polarization $P_0=1-\sigma_{\text{off}}/\sigma_{\text{on}}$ in a ribbon with periodic boundary conditions vs. length $L$ (in units of the lattice constant $a_0$). The inset shows the on- and off-conductances, $\sigma_{\text{off}}$ and $\sigma_{\text{on}}$, respectively. Here, $e A_1 a_0/c=1.0, A_4/A_1=0.1,\Omega/\gamma\approx 1.47$, where $\gamma\approx2.7$~eV is the hopping parameter, and the ribbon width $w=6\sqrt{3}a_0$.}\label{fig:graphene}
\end{figure}

In this work we discuss a fundamentally different approach to
optically-controlled valleytronics in the bulk that offers a high degree of tunability in a single sample.
In this approach, the light is relatively
intense, so that the electronic structure is represented by eigenvalues
of a Floquet Hamiltonian.  The time-dependence of the
electric field, rather than intrinsic properties of the material
or nanoscale structures, is used to effectively break inversion symmetry and
distinguish the valleys.  One way to do this, for example in graphene, is by shining an admixture
of circularly polarized light of frequencies $\Omega$ and $4\Omega$, which
can be coherently generated by use of nonlinear crystals.  As explained
below, for appropriate choices of amplitudes and
phase offset one can produce a Floquet quasienergy spectrum which is
gapped for one valley but gapless for the other.  A dc current passed
between leads with chemical potentials in this gap is then valley-polarized.
The degree of polarization can be interrogated with different phase-offsets
in the vicinities of each lead, such that the gap closing is in the same or different
valleys for each.  An example of this behavior for an idealized system is
illustrated in Fig.~\ref{fig:graphene}.  The polarization
turns out to be quite robust against disorder and edge effects, as shown
below.  This system represents an optically controlled valley valve.

This physics can also be applied to systems in which inversion symmetry is already
broken, such as graphene on a BN substrate~\cite{gorbachev_2014} or dichalcogenide
materials like MoS$_2$ and WS$_2$, which have pre-existing gaps that are the same
for both valleys.  Circularly polarized light
may close one gap while opening the other in such materials, again allowing marked preferential
conduction for one of the two valleys.  Left- and right-circularly polarized
light create open channels for opposite valleys, leading to optically-controlled
valley polarization, as we demonstrate below.

\emph{Optically Broken Inversion Symmetry in Graphene}---%
In the presence of a temporally periodic potential, electronic states
follow the time-dependent Schr\"odinger equation.
Floquet's theorem~\cite{rahzavy_2003} guarantees that
its solutions as a function of time $t$ have the form
$
\psi_{\alpha}(t)= u_{\alpha}(t)
e^{- i\varepsilon_{\alpha} t},
$
with $u_{\alpha}(t+T)=u_{\alpha}(t)$,
where $T$ is the period of the Hamiltonian $H(t)$, and $\alpha$
includes any quantum numbers required to specify the electronic state.
The quasienergies $\varepsilon_{\alpha}$
(which may be restricted to the interval $-\Omega/2 < \varepsilon_{\alpha} \le \Omega/2$)
are eigenvalues of the ``Floquet Hamiltonian''
$H_F(t)=H(t)-i\partial_t$, and $u_{\alpha}$ are the corresponding eigenfunctions.

%----- Fig. 2 -----%
\begin{figure}[t]
\includegraphics[width=3.4in]{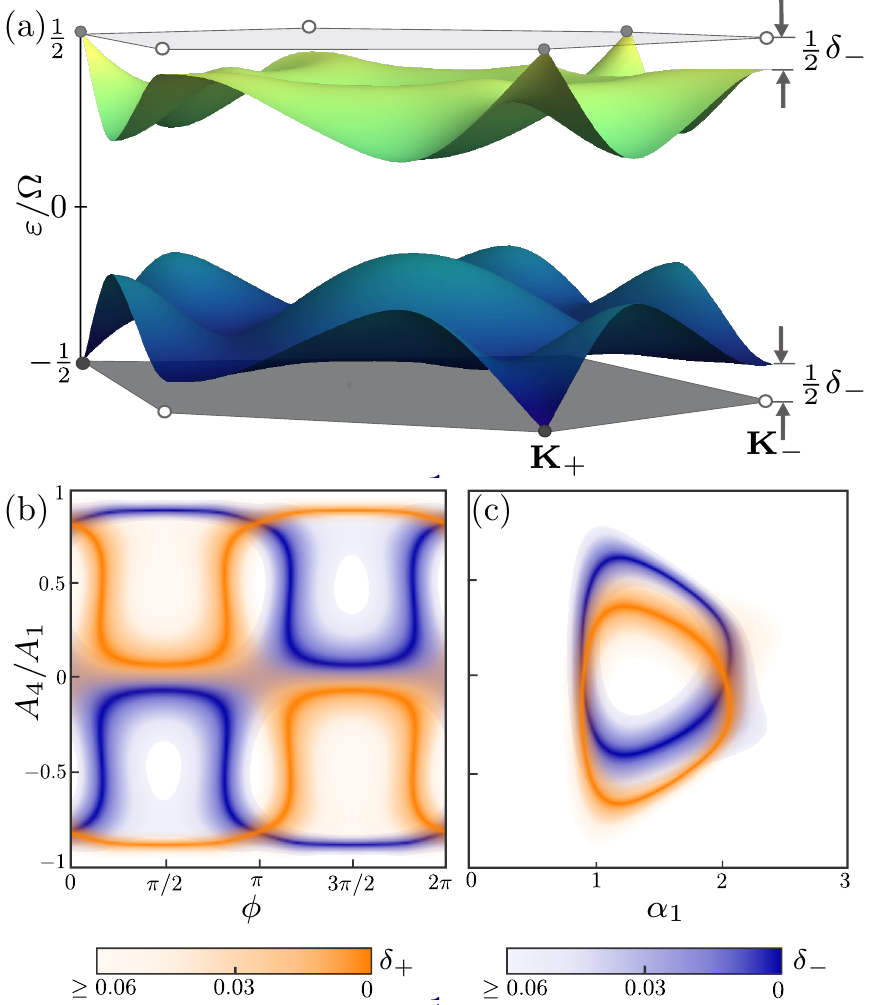}
\caption{(color online) (a) The bulk quasienergy spectrum in the Brillouin-Floquet zone for uniformly irradiated graphene ($eA_1a_0/c=1.5, A_4/A_1=0.6, \phi=\pi/2, \Omega/\gamma=1.44$) shows a gap at $\Km$ and gapless Dirac dispersion at $\Kp$ point. (b) and (c) The quasienergy gaps $\Omega\delta_\pm$ of the $\Kp$ (blue) and $\Km$ (orange) valleys at $\varepsilon=\Omega/2$ show the gaps may be opened and closed by tuning the optical parameters. In (b) $\alpha_1\equiv eA_1a_0/c=1.5, \Omega/\gamma=1$ and in (c) $\phi=0, \Omega/\gamma=1.44$.}\label{fig:bulk}
\end{figure}

As a paradigm of these systems we consider the case of graphene, in which the
electrons are assumed to hop around on a tight-binding honeycomb lattice
~\cite{Castro_Neto_RMP,peres_2010,dassarma_2011}, and the effects of
the circularly polarized light are implemented via time-dependent phases in
the hopping matrix elements~\cite{oka_2009}. Variations in the intensity
and the frequency of the light tune the quasienergy band structure through many distinct topological phases~\cite{kundu_2014}.
The energy spectrum
includes two valleys of states near the $\Kp$ and $\Km$ points
of the Brillouin zone, and wavefunctions are two-component spinors
representing the electron amplitudes on the two sublattices of the
honeycomb structure.  In the presence of a spatially uniform, time-dependent
electric field from light normally incident on the graphene plane, the Floquet Hamiltonian in the sublattice basis $(u_A, u_B)$ has the form~\cite{kundu_2014}
%%%%
\begin{equation}\label{eq:ham}
H_F({\bf k},t)=
\left(
\begin{array}{cc}
-i\partial_t & -\gamma Z({\bf k},t)  \\
-\gamma Z^*({\bf k},t) & -i\partial_t
\end{array}
\right),
\end{equation}
%%%%
where ${\bf k}$ is the wavevector of the state, $\gamma$ is a hopping amplitude,
$Z({\bf k},t)=\sum_{n=1}^3
e^{i[{\bf k}+\frac{e}{c} {\bf A}(t)] \cdot {\bf a}_n }$,
${\bf a}_n$
are the nearest neighbor vectors of a site on the lattice,
and ${\bf A}(t)=A_1(\cos\Omega t,\sin\Omega t)$
is the vector potential for the electric field, with $\Omega=2\pi/T$.
%\highlight{The laser is focused on an area larger than typical system sizes; thus, we may ignore the wavelength of $\vex A$}.
When
${\bf k}$ is set to the $\Kpm$ point [$\Kpm
= (0,\pm4\pi/3\sqrt{3}a_0)$,
with $a_0$ the nearest neighbor distance] the lattice symmetry
combines with the temporal symmetry so that
$Z(\Kpm,t+T/3)=e^{\mp2\pi i/3}Z(\Kpm,t)$, effectively
tripling the relevant frequency for the Floquet problem.
Because of this,
distinct states can cross, rather than repel, at the Floquet zone boundary
$\varepsilon_{\alpha}= \pm\Omega/2$ as $A_1$ or
$\Omega$ are varied, leading to topological transitions in the quasienergy
band structure.

In this situation such crossings occur at the $\Kpm$ points simultaneously.
This results from
a combination of inversion symmetry ($\Km = -\Kp$)
and a form of time reversal symmetry: If
$u(\Kp,t)$ is an eigenvector of $H_F(\Kp,t)$, then
$\sigma_x u^*(\Kp,-t)$, with $\sigma_x$ a Pauli matrix,
is an eigenvector of $H_F(\Km,t)$ with the same eigenvalue.
Lifting this coincidence
of eigenvalues distinguishes the valleys.  One way to do this is by
changing $H_F$ such that
%%%%
\begin{equation}\label{eq:sym_break}
Z(\Kp,-t) \neq Z(\Km,t),
\end{equation}
%%%%
while retaining the $T/3$ period of $e^{\pm i\Omega t}Z(\Kpm,t)$
needed for a gap closing,
breaking the effective inversion symmetry.  In contrast to approaches
which use static potentials ~\cite{song_2013,grujic_2014},
here we seek to use the vector potential to do so,
allowing for optical control of the valley-distinguishing properties of the system.

One way this can be done is by adding a 4th harmonic to the vector potential,
so that it has the
form $A_x+iA_y = A_1 e^{i\Omega t} + A_4 e^{i(4\Omega t + \phi)}$.
For this $\vex A$ the inequality~(\ref{eq:sym_break}) is satisfied, provided $\phi \ne 0,\pi$, while
the frequency tripling of $e^{\pm i\Omega t}Z(\Kpm,t)$ is retained.
As illustrated in Fig.~\ref{fig:bulk}, gaps at $\Kpm$ are in general unequal and can be opened and closed around $\varepsilon=\Omega/2$ separately by tuning the optical parameters.
Note that the phase offset $\phi$ may be
adjusted by varying the optical path length of the $4\Omega$ light component
relative to the $\Omega$ component. As we next show numerically,
this leads to different band gaps for the two valleys, even allowing
the gap to close for one while the other remains open.  Thus, the admixture of the
two frequencies of light in principle allows one to prohibit a bulk current
for one valley while allowing it for the other.

\emph{Numerical Results.}---%
To test this idea, we have computed the DC conductance at zero temperature for an irradiated graphene strip~\cite{gu_2011,kundu_2014}.
Armchair graphene ribbons were simulated, with periodic boundary conditions across the width to minimize the edge effects while keeping the size of the system small enough for numerical efficiency. \highlight{Optical parameters were chosen to produce quasienergy gaps large enough to observe the effect with our simulated system sizes. More realistic optical parameters are discussed below.} %The leads were modeled as highly doped (to $1/6$th of their bandwidths) semi-infinite graphene ribbons.
We also performed simulations for more realistic systems with open boundary conditions. These results are reported below and support our conclusions (see Appenix~\ref{app:B}). %We used the Floquet Green's function formalism~\cite{arrachea2005} to calculate the conductance~\cite{suppl} and employed the recursive Green's function method~\cite{Sancho} for the computation of the density of states. The leads were treated in the wide-band approximation~\cite{kundu_2014}.

The current is introduced from the leads attached to the two ends of the system, while the central graphene region is voltage-biased to align the Floquet zone edge with the average chemical potential of the leads. In this scheme, the entire system is illuminated by light of frequency $\Omega$.  Each half of the system around the leads is further illuminated by circularly-polarized light of frequency $4\Omega$ with independently controlled values of the phase offset, $\phi_L$ and $\phi_R$. With $\phi_L=\phi_R$, due to the gap for one valley, the system essentially allows only current from the other valley to pass, yielding
a valley-polarized current.
That this current is valley-polarized is confirmed by comparing with the conductance when $\phi_L-\phi_R= (2n+1)\pi$, with $n$ an integer. In this case, the two halves are conductive for opposite valleys so very little net current passes.  This means the system behaves
as a valley valve~\cite{rycerz_2007b} which may be opened or closed optically. The relative
conductance in these two cases ($\sigma_{\text{on}}$ and $\sigma_{\text{off}}$, respectively)
offer a measure of the valley polarization, $P_0 \equiv 1-\sigma_{\text{off}}/\sigma_{\text{on}}$,
achieved in the system. The fidelity of the resulting valve can exceed $P_0=98$\%, as illustrated in Fig.~\ref{fig:graphene}(b) (see also Appendix~\ref{app:D}. %\highlight{(The length dependence of $P_0$ is discussed in the Supplemental Material~\cite{suppl}.)}
This is one of our main results.

%----- Fig. 3 -----%
\begin{figure}[t]
\centering
\includegraphics[width=3.4in]{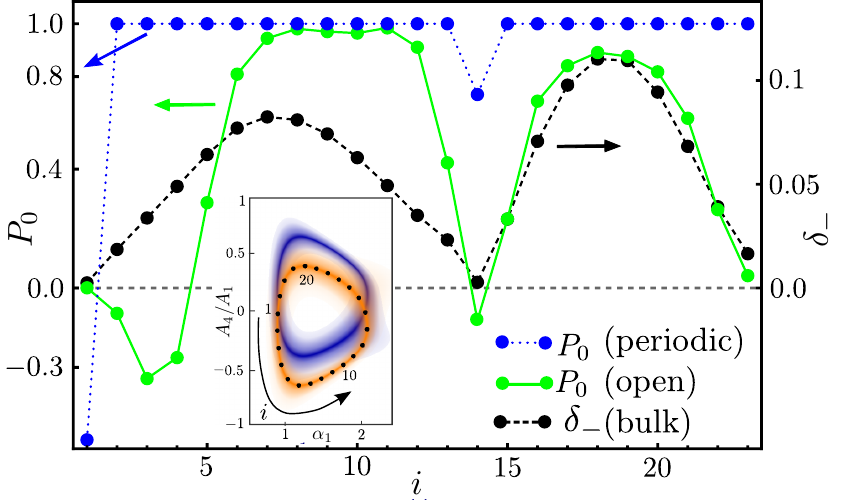}
\caption{(color online) The valley polarization $P_0$ for the system with periodic (dotted blue line) and open (solid green line) boundary conditions, respectively. The index $i$ indicates parameter values at points along the $\delta_+=0$ (orange) contour shown in the inset, as in Fig.~\ref{fig:bulk}(c). The corresponding gap $\delta_-$ is also shown (dashed black line). The fixed parameters are as in Fig.~\ref{fig:bulk}(a); the chemical potential of the irradiated region is $0.71\gamma\approx\Omega/2$; the periodic system has a full length $L=2\ell=288a_0$
and width $w=6\sqrt{3}a_0$; the open system has $\ell=27a_0$, $w=24\sqrt3a_0$; the leads have a width $12\sqrt3a_0$ and are connected equidistant from the edges across the width and a distance $9a_0$ away from the center.}\label{fig:edges}
\end{figure}

\emph{Effects of Edges.}---%
The valley polarization in our proposal results from bulk transport. Due to the existence of multiple Floquet topological phase transitions tuned by frequency~\cite{kundu_2014}, the system with open boundary conditions has, in addition, a number of chiral edge states. \highlight{These edge states can carry current but lack a well-defined valley index. Moreover, they can scatter the bulk states between valleys. Both of these effects can degrade the observed valley polarization, and are particularly noticeable in small systems. We expect that in sufficiently large sample the degrading effects of the edges will be much reduced.}

To support this expectation, we also simulated a more realistic system with open boundary conditions. \highlight{To minimize edge effects, we employ a geometry in which the leads
are connected away from them, as illustrated in Fig.~\ref{fig:graphene}(a).
This requires relatively large widths, limiting the system lengths one can ultimately
study efficiently.  Nevertheless, one may still obtain information about the large-length limit by choosing a quasienergy gap that is not too small.}
We report our results in Fig.~\ref{fig:edges} and compare to the case with periodic boundary conditions. \highlight{The path followed in Fig.~\ref{fig:edges}(a) is chosen so that the gap for one of the valleys vanishes precisely, maximizing the ``on'' current and thereby the valley polarization. The system with periodic boundary conditions and a gapless $\Kp$ point shows nearly 100\% valley polarization for an arbitrary value of the quasienergy gap $\delta_- >0$ at $\Km$. For the system with open boundary conditions, a larger gap $\delta_-$ is required to have significant valley polarization. Even so, remarkably large valley polarizations are obtained in this case for a relatively small system (see also Appendix~\ref{app:B}).} %In Fig.~\ref{fig:edges}(b) we show the polarization $P(\Delta\phi)\equiv1-\sigma(\pi)/\sigma(\Delta\phi)$ as a function of $\Delta\phi\equiv\phi_R-\phi_L$ for both periodic and open boundary conditions. In either case, the system shows almost purely valley-polarized transport and a large ``on'' signal for a wide range around $\Delta\phi=0$.} %(Note that $P_0=P(0)$.)

%----- Fig. 4 -----%
\begin{figure}[t]
\centering
\includegraphics[width=3.4in]{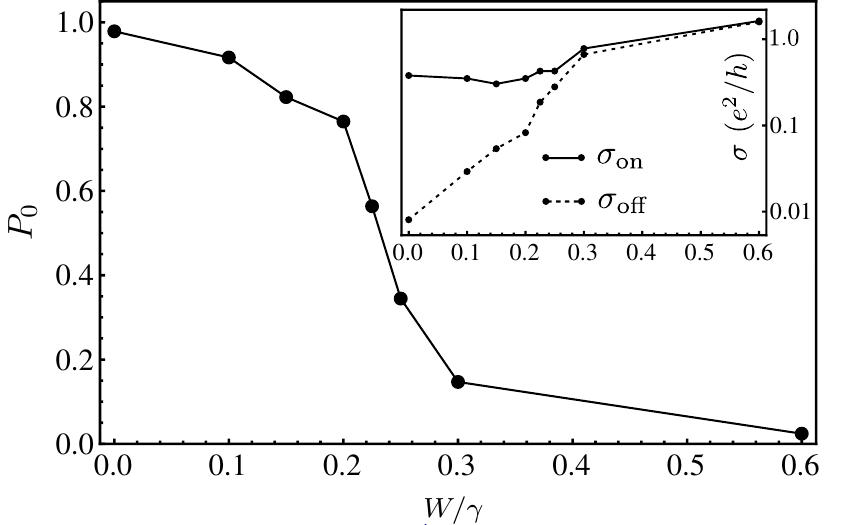}
\caption{The polarization of the graphene valley valve vs. the strength of onsite static disorder potential, $W$. The parameters are as in Fig.~\ref{fig:bulk}(a) and the chemical potential of the irradiated region is $0.71\gamma\approx\Omega/2$. The boundary conditions are open and the dimensions are as for the open system in Fig.~\ref{fig:edges}.}\label{fig:disorder}
\end{figure}

\emph{Dirac Systems with Statically Broken Inversion Symmetry.}---%
A second way to break the symmetry of the valley electronic states and realize an optically-controlled valleytronic system
is by statically lifting the inversion symmetry of the system. Examples of such systems are provided by
graphene deposited on boron nitride~\cite{HunSanYou13a,CheShiYan14a} and single-layer  MoS$_2$~\cite{mak_2010,mak_2014}.
A simple description of these systems is a tight-binding
model on a honeycomb lattice with a staggered potential $\pm\mu_s$ of
opposite signs on different sublattices; the low energy electron states
at the $\Kpm$ points are then governed by massive Dirac Hamiltonians. Despite this apparent inversion symmetry breaking, the magnitude of the gap in both valleys is the same and no valley polarization can be realized in equilibrium.
Valley polarization in a static region can be achieved in this
system via optical absorption~\cite{zeng_2012,mak_2012}.

In our approach, the gaps at the $\Kpm$ points
can be distinguished by the helicity of the monochromatic circularly-polarized light. The Floquet Hamiltonian for the low-energy excitations now takes the form $H_F({\bf k},t)+M$, where $H_F$ is the same as in Eq.~(\ref{eq:ham}), and
\begin{equation}
M = \left(
\begin{array}{cc}
\mu_s & 0  \\
0 & -\mu_s
\end{array}
\right).
\end{equation}
Due to the inversion-symmetry breaking, the quasienergies around the two valleys now evolve differently: for a given helicity, the frequency can be tuned to a value 
such that there is a
gapless $\Kp$ valley crossing $\varepsilon=0$, and a gapped $\Km$ valley (see Appendix~\ref{app:C}). Illuminating the two halves of the system with circularly-polarized lasers at such a frequency realizes a valley valve. The valve can be turned on and off by switching the helicity of one of the lasers. We have checked that, in this case, the current induced from the leads at energy $E\approx0$ has
valley polarization exceeding 90\%.

\emph{Discussion.}---%
\highlight{To observe the effects we have demonstrated above, energy and length scales in real samples must be chosen appropriately. For example, the temperature in the leads must be less than the quasienergy gap of the gapped valley, namely, $\Omega\delta_-$. This can be tuned optically. For example, at the main frequency $\Omega=0.2\gamma=0.54$~eV, a gapless $\Kp$ is obtained at $\alpha_1=eA_1a_0/c=0.1$ and laser intensity $I_1=(1/4\pi c)A_1^2\Omega^2\sim10^{14}$~W$/$m$^2$. %~\cite{Note1}.
At these values, by varying the ratio $A_4/A_1$, the gap at $\Km$ can be tuned to be $>5$~meV. Increasing the frequency and the intensity of the laser can produce even larger gaps (see Appendix~\ref{app:A}). Since the conductance of the gapped valley decays over the wavefunction evanescent length, $\ell_{\text{ev}}\propto v/\Omega\delta_-$, the length of the system $L$ must exceed $\ell_\text{ev}$. Here, $v$ is the Fermi velocity at the valley, which can also be tuned optically. In our estimate, for the aforementioned values, $\ell_\text{ev}<500a_0=71$~nm}

\highlight{In practice, there are always extrinsic perturbations limiting the polarization. Prominent among these is disorder, which does so through inter-valley scattering. Intuitively, the effect of disorder is to fill in the gap in the quasienergy spectrum. We expect that its effect is controlled by this gap, and that it should not spoil the valley polarization when weak enough. In Fig.~\ref{fig:disorder} we show results of our simulations of disorder-averaged polarizations in the graphene system discussed above (see also Appendix~\ref{app:C}). In these simulations we have used random, static onsite potentials, with Gaussian distribution of zero mean and standard deviation $W$ characterizing the disorder strength. The conductances are each averaged over 25 disorder configurations. Indeed, at larger strengths, disorder causes intervalley scattering and diminishes the valley polarization by increasing $\sigma_{\text{off}}$ (see inset). However, when $W$ is  sufficiently smaller than the quasienergy gap, here $\Omega\delta_- \approx 0.1\gamma$, the valley polarization we obtain is robust against disorder, exceeding 70\% over a significant range of disorder strength. As discussed above, in real samples the quasienergy gap can be tuned to be $>5$~meV. Samples with disorder strengths below this appear to be currently available in the lab~\cite{Du2008,Bolotin2008,Xue2011,Efetov2015}. Thus we believe there is considerable room to vary the parameters of the system without having the effects spoiled by disorder.}

Novel forms of electronic states arise as the result of the interaction between electrons and periodic external driving fields. The optically controlled valley polarization and valleytronic devices proposed in this work promise new ways of engineering and utilizing the emergent electronic degrees of freedom in graphene and related Dirac systems.

This work was supported in part by the NSF through
Grant Nos. DMR-1350663 and DMR-1506460, the US-Israel
Binational Science Foundation, and by Indiana University.

\appendix

\section{Optical parameters}\label{app:A}

\begin{figure}[ht]
\includegraphics[width=3in]{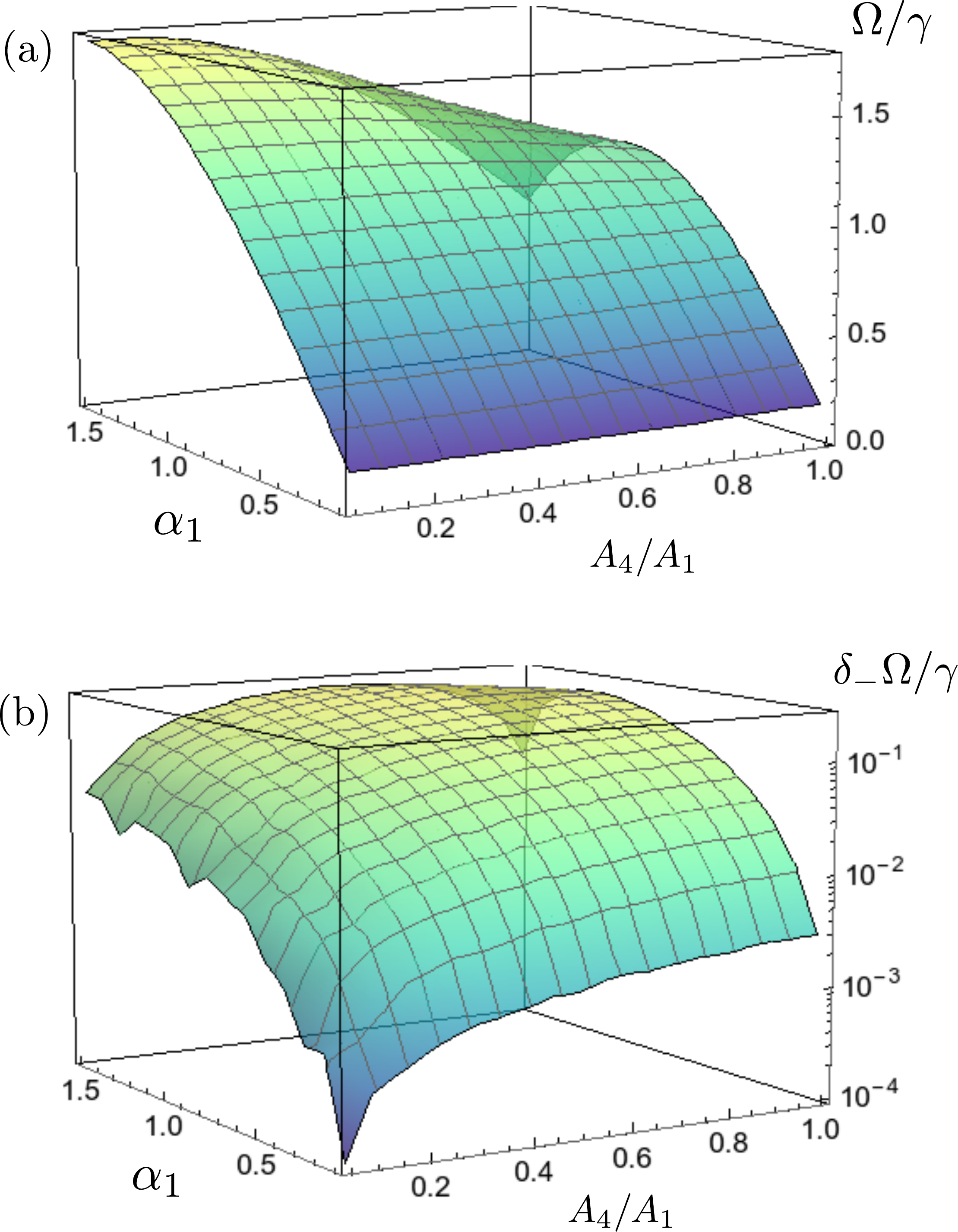}
\caption{(a) The optical parameters $\alpha_1=eA_1a_0/c$, $A_4/A_1$ and the largest $\Omega/\gamma$ at which the quasienergy gap $\delta_+=0$ at quasienergy $\varepsilon=\Omega/2$ for the graphene device. (b) The corresponding gap at $\mathbf{K_-}$.}\label{fig:S1}
\end{figure}

  Here we first show numerical results for the optical parameters needed for the graphene valley valve discussed in the main text and in Fig.~1. In Fig.~\ref{fig:S1}(a), we plot the solutions to $\delta_+(A_1,A_4,\Omega)=0$ with the largest $\Omega$, i.e. the points at which the quasienergy gap at $\varepsilon=\Omega/2$ first vanishes for the $\Kp$ valley. In Fig.~\ref{fig:S1}(b) we plot the value of the gap at these points for $\Km$ point. Together, these plots can be used to guide the choice of optical parameters in an experiment. For example, laser intensity of the main frequency $I_1=(1/4\pi c)A_1^2\Omega^2\sim10^{14}$~W$/$m$^2$ for $\alpha_1=eA_1a_0/c=0.1$ and $\Omega=0.2\gamma=0.54$~eV, which is within current capabilities. By adjusting the intensity of the 4th harmonic, one can tune the gap for the $\Km$ point shown in Fig.~\ref{fig:S1}(b) by more than an order of magnitude; for an intensity $I_4\sim I_1$ in graphene ($\gamma\approx 2.7$~eV), this gap is $\sim5$~meV.
  
  %----- Fig 5 ------%
  
\section{Numerical methods}\label{app:B}
We have employed the Green's function method~\cite{arrachea2005} for computing the conductance of the system numerically. The recursive Green's function method~\cite{Sancho} is employed for the computation of the density of states. The leads were modeled as highly doped (to $1/6$th of their bandwidths) semi-infinite graphene ribbons.

%----- Fig 6 ------%
\begin{figure}[tb]
\centering
\includegraphics{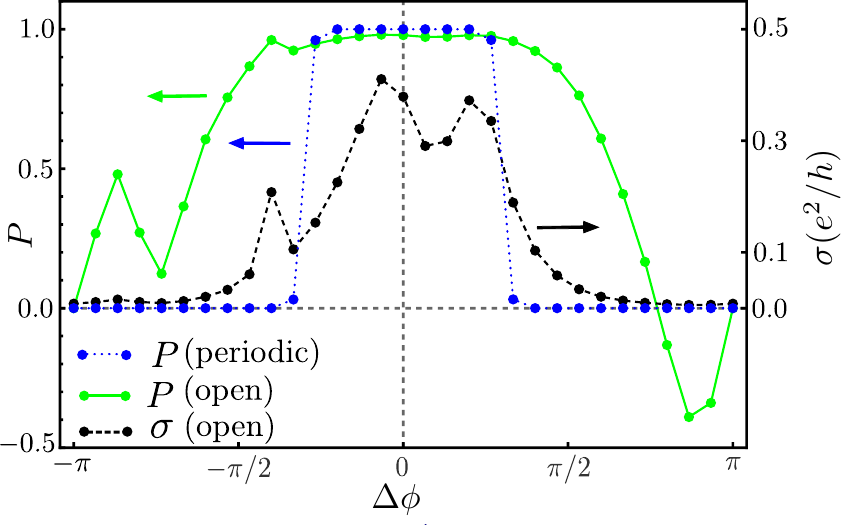}
\caption{The polarization $P(\Delta\phi)\equiv1-\sigma(\pi)/\sigma(\Delta\phi)$ for the graphene system with periodic (dotted blue line) and open (solid green line) boundary conditions, respectively,  vs. $\Delta\phi\equiv\phi_R-\phi_L$. Also shown is the conductance $\sigma(\Delta\phi)$ at zero temperature (dashed black line) for the system with open boundary conditions. The fixed parameters are as in Fig.~2(a) of the main text; the chemical potential of the irradiated region is $0.71\gamma\approx\Omega/2$; the periodic system has a full length $L=2\ell=288a_0$
and width $w=6\sqrt{3}a_0$; the open system has $\ell=27a_0$, $w=24\sqrt3a_0$; the leads have a width $12\sqrt3a_0$ and are connected equidistant from the edges across the width and a distance $9a_0$ away from the center.}\label{fig:S2}
\end{figure}

%----- Fig 7 ------%
\begin{figure*}[bt]
\centering
\includegraphics[width=0.95\textwidth]{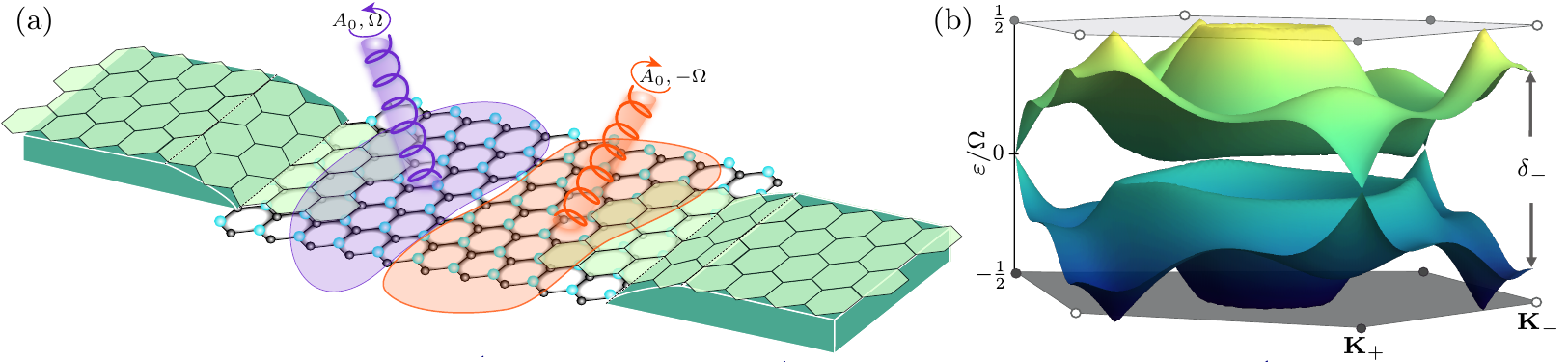}
\caption{(a) Schematic device with two circularly polarized light of opposite helicity incident on a Dirac system with broken inversion symmetry (such as MoS$_2$). (b) The quasienergy spectrum for a uniformly irradiated system modeled with a staggered chemical potential with fixed helicity  shows a gapless $\Kp$ point at $\varepsilon=0$ and a gapped $\Km$ point; The gapped and gapless points are switched for the opposite helicity. The staggered chemical potential $\mu_s=0.4\gamma$. The other parameters $\alpha_1 = eA_1a_0/c = 0.5,\Omega/\gamma = 1.73$, and the chemical potential of the irradiated region is $0.12\gamma$.}\label{fig:S3}
\end{figure*}

%----- Fig 8 ------%
\begin{figure}[t]
\centering
\includegraphics{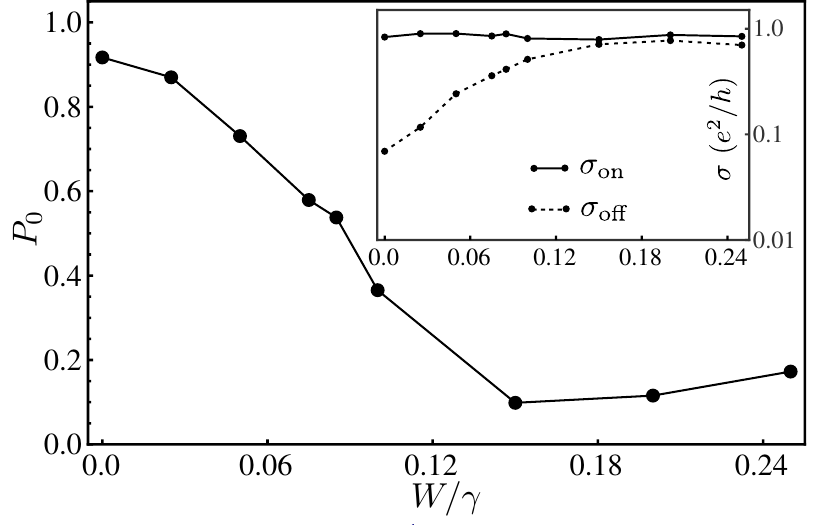}
\caption{The polarization of a gapped Dirac system  vs. the strength of onsite static disorder potential, $W$. The parameters used are as in Fig.~\ref{fig:S3}, but 
the system has open boundary conditions. The dimensions of the system are $w=15\sqrt3a_0,\ell=54a_0$, the leads have a width $7\sqrt3a_0$ and connected a distance $36a_0$ away from the center.}\label{fig:S4}
\end{figure}

The system is described by the Hamiltonian $H_S(t)=H(t)+H_C$, where $H(t) = H(t+T)$ is the Hamiltonian of the irradiated area,
%(Time period $T$ is related to the driving frequency $\Omega$ by $T=2\pi/\Omega$),%
and the contact Hamiltonian is
%%%%
$$
H_C = \sum_{\lambda=L,R} F^{\lambda}_{r l} a^{\lambda\dagger}_{r}c_l +~\text{h.c.}
\equiv \sum_{\lambda=L,R} a^{\lambda\dagger}F^{\lambda}c +~\text{h.c.},
$$
%%%%
where $L,R$ denote left and right leads. The $a^{\lambda\dagger}_{r}$ and $c^{\dagger}_{l}$
denote, respectively, the electronic creation operator at site $r$ for lead $\lambda$ and at site $l$ for the irradiated area. The time dependent current through lead $\lambda$ is $J^{\lambda}(t) = ie\left[H(t) +H_C, N^{\lambda}(t) \right]$, with $N^{\lambda}$ the number operator. The time averaged current in the left lead is
$I=\int_0^T\langle J^{L}(t)\rangle\dd t/T$, where $\langle \cdot\rangle$ denotes an average over the lead states. It may be written in the form~\cite{kohler2005}
%%%%
$$
I=\frac{e}{2\pi}\int \dd\omega \sum_{k\in \mathbb{Z}}\left[ T^{(k)}_{LR}(\omega)f_R(\omega) - T^{(k)}_{RL}(\omega)f_L(\omega)\right],
$$
%%%%
where $f_{\lambda}(\omega) = [1+e^{(\omega - eV_{\lambda})/\tau_{\lambda}}]^{-1}$ is the Fermi-Dirac distribution in lead $\lambda$, with electric bias $V_{\lambda}$ and temperature $\tau_{\lambda}$. (We adopt units in which the Boltzmann constant $k_B=1$.)
The transmission probability $T^{(k)}_{\lambda \lambda'}$ is given by
%%%%
\begin{align}
T^{(k)}_{\lambda \lambda'} = \text{Tr}\left[G^{(k)\dagger}(\omega)\xi^{\lambda}(\omega+k\Omega)G^{(k)}(\omega)\xi^{\lambda'}(\omega) \right],\nonumber
\end{align}
%%%%
where the Floquet Green's function is
$$
G^{(k)}(\omega)=
\frac1T\int_0^T dt\int ds\: G(t,t-s)e^{i\omega s}e^{ik\Omega t}, \quad k\in\mathbb{Z}
$$
in terms of the time-dependent Green's function of the system $G(t,t')=G(t+T,t'+T)$.
%\begin{align}
%G(t,t') = \sum_{k\in\mathbb{Z}}\frac{d\omega}{2\pi} G^{(k)}(\omega)e^{-i\omega(t-t')}e^{-ik\Omega t},\nonumber
%\end{align}

The coupling between the irradiated area and the leads enters through the factors $\xi^{\lambda}(\omega)=2\pi F^{\lambda\dagger}\rho^{\lambda}(\omega)F^{\lambda}$, where $\rho^{\lambda}(\omega)$ is the density of states in lead $\lambda$. Finally, for zero relative bias of the leads, the zero-temperature conductance is given by
$$
\sigma = \frac{e^2}{2\pi} \sum_{k\in\mathbb{Z}}\left[ T^{(k)}_{LR}(0) + T^{(k)}_{RL}(0)\right].
$$
The conductance computation is simplified by the use of the ``wide band limit'', $\rho^{\lambda}(\omega) = \rho^{\lambda}$ independent of $\omega$~\cite{kundu_2014}. In Fig.~\ref{fig:S2}, we show the results of the simulation for conductance and valley polarization as a function of the phase difference $\Delta\phi=\phi_R-\phi_L$, as in the setup shown in Fig.~1(a) of the main text, for both periodic and open boundary conditions. In either case, the system shows almost purely valley-polarized transport and a large ``on'' signal for a wide range around $\Delta\phi=0$.

%----- Fig 9 ------%
\begin{figure}[t]
\centering
\includegraphics[width=3.3in]{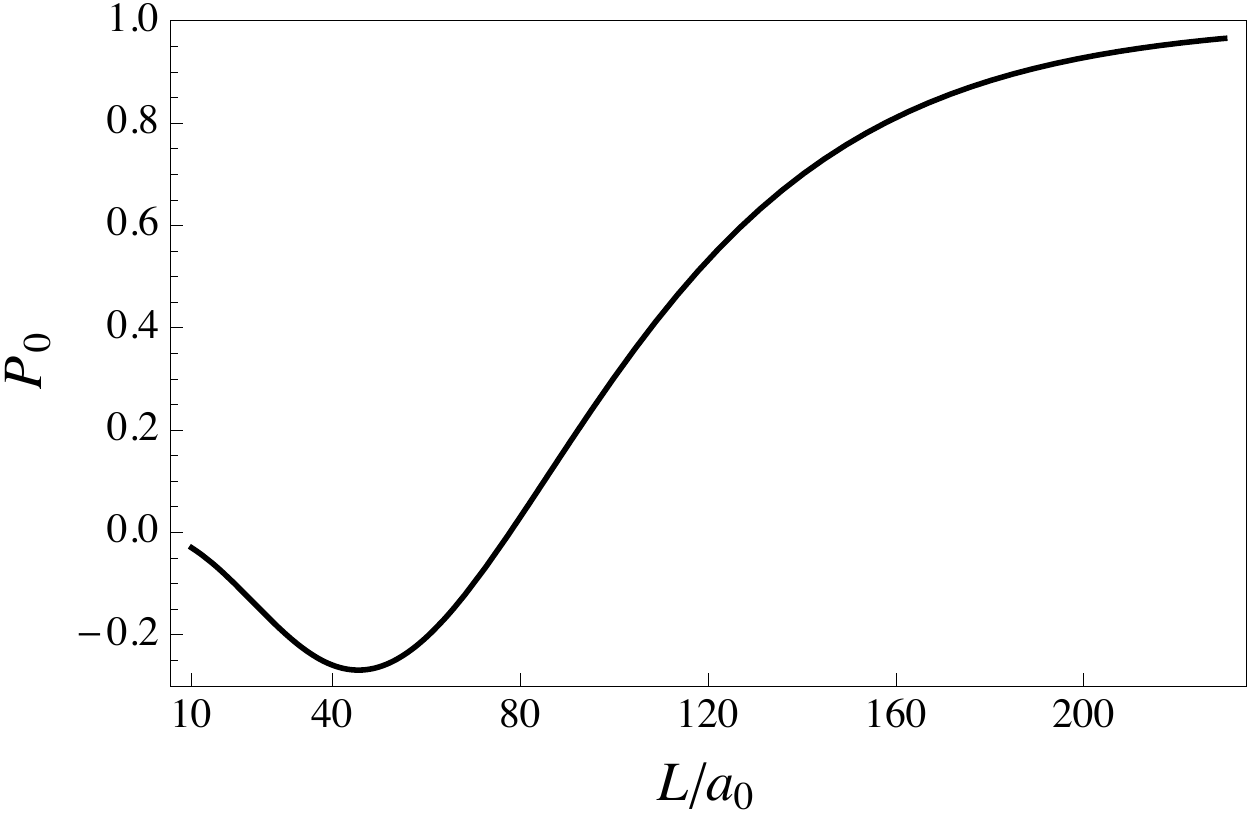}
\caption{Modeling the valley polarization vs. length $L$ of the system shown in Fig.~1(a). The width $w=12a_0$, the evanescent length $\ell_\text{ev}=20a_0$ and the interpolating function $f(x)=\tanh x$.}\label{fig:S3}
\end{figure}

\section{Dirac system with broken inversion symmetry}\label{app:C}
A schematic of the system to create valley-polarized current using statically broken Dirac system is shown in Fig.~\ref{fig:S3}(a). A typical quasienergy spectrum is shown in Fig.~\ref{fig:S3}(b), where the inversion symmetry is broken by staggered potential, as in Eq.~(3) of the main text. Also, in comparison with Fig. 4 of the main text, we show in Fig.~\ref{fig:S4} the effect of disorder on the valley polarization in this case.

\section{Length dependence of the valley polarization}\label{app:D}
As seen in Fig.~1(b), the valley polarization, $P_0=1-\sigma_\text{off}/\sigma_\text{on}$, is negative for a range of lengths before approaching 1 for larger lengths. In this section, we present a model for this length dependence.

The reason for this particular behavior lies in the interplay of diffusive and evanescent transport in a finite size system.
As a simple model, we shall assume that the two valleys behave as two independent conduction channels in each half of the device shown in Fig.~1(a). 
The gap at one of the valleys, say $\Kp$, vanishes at large system sizes (diffusive conduction channel) 
while the other one, say $\Km$, remains finite (evanescent conduction channel). 
In the ``off'' configuration, the valley currents are carried in parallel within each half of the system, and the two halves are connected in series.
In the ``on'' configuration, the valley currents are carried in parallel along the whole length of the system. At small system sizes, the evanescent length  becomes of order $L$, 
the distance between the leads. In this regime, both channels are diffusive. As the system size increases, only one of the parallel channels in the ``on''
configuration remains diffusive; in the ``off'' configuration, both parallel channels behave similarly. This causes a 
faster drop in the ``on'' conductance in intermediate length scales. At yet larger length scales,
the ``on'' configurations remains diffusive in one channel, while the ``off'' current
becomes completely evanescent and decays exponentially. 

To make this more precise, we take the conductance $\sigma$ for each valley to be
%%%%%
\begin{equation}
\sigma(l,\xi)=\frac{w}{l}\exp\left[-\frac{l}{\xi}\right].
\end{equation}
%%%%%
Here, $w$ is the width of the system, $l$ is the length of the channel, and $\xi$ is a length scale inversely proportional to the valley's quasienergy gap. 
For the $\Kp$ valley, transport is diffusive and we take $\xi_+=L$.  For the $\Km$ valley, there is a crossover from diffusive to evanescent transport; so, we take $\xi_-=\ell_\text{ev}f(L/\ell_\text{ev})$, where $\ell_\text{ev}$ is the evanescent length and $f(x)$ is an interpolating function satisfying $f(x\ll1)=x$ and $f(x\gg1)=1$. The ``on'' and ``off'' conductances are given by
%%%%%
\begin{align}
\sigma_\text{on} &=\sigma(L,\xi_+)+\sigma(L,\xi_-), \\
\sigma_\text{off} &=2 \left[ \frac1{\sigma(L/2,\xi_+)}+\frac1{\sigma(L/2,\xi_-)} \right]^{-1}.
\end{align}
%%%%%
Our results are plotted in Fig.~\ref{fig:S3} for a representative set of parameters, showing the same behavior for $P_0(L)$ in this model as in Fig.~1(b).

%\bibliographystyle{physre}
%\bibliography{mia_2015}

\vspace{-2mm}
\begin{thebibliography}{10}
\vspace{-2mm}

\bibitem{Novoselov_2004}
K.S. Novoselov et al., Science {\bf 306}, 666 (2004).

\bibitem{Castro_Neto_RMP}
A.~H. {Castro Neto}, F.~Guinea, N.~M.~R. Peres, K.~S. Novoselov, and A.~K.
  Geim,
\newblock \rmp\ {\bf 81}, 109 (2009).

\bibitem{rycerz_2007b}
A.~Rycerz, J.~Tworzydlo, and C.~Beenakker,
\newblock Nature Phys. {\bf 3}, 172 (2007).

\bibitem{song_2013}
Y.~Song, F.~Zhai, and Y.~Guo,
\newblock Appl. Phys. Lett. {\bf 103}, 183111 (2013).

\bibitem{grujic_2014}
M.~M. Grujic, M.~Z. Tadic, and F.~M. Peeters,
\newblock \prl\ {\bf 113}, 046601 (2014).

\bibitem{niu_2008}
W.~Yao, D.~Xiao, and Q.~Niu,
\newblock \prb\ {\bf 77}, 235406 (2008).

\bibitem{abergel_2009}
D.~Abergel and T.~Chakraborty,
\newblock Appl. Phys. Lett. {\bf 95}, 062107 (2009).

\bibitem{mak_2014}
K.~Mak, K.~McGill, J.~Park, and P.~McEuen,
\newblock Science {\bf 344}, 1489 (2014).

\bibitem{gorbachev_2014}
R.~Gorbachev {\em et~al.},
\newblock Science {\bf 346}, 448 (2014).

\bibitem{qi2014}
F.~Qi and G.~Jin,
\newblock J. Appl. Phys. {\bf 115}, 173701 (2014).

\bibitem{SieMcILee15a}
Edbert J. Sie, James W. McIver, Yi-Hsien Lee, Liang Fu, Jing Kong, and Nuh Gedik,
\newblock Nature Mat. {\bf 14}, 290 (2015).

\bibitem{cao_2012}
T.~Cao {\em et~al.},
\newblock Nature Comm. {\bf 3}, 887 (2012).

\bibitem{zeng_2012}
H.~Zeng, J.~Dai, W.~Yao, and X.~Cui,
\newblock Nature Nanotech. {\bf 7}, 490 (2012).

\bibitem{shan_2015}
W.~Y. Shan, J.~Zhou, and D.~Xiao,
\newblock \prb\ {\bf 91}, 035402 (2015).

\bibitem{lensky_2014}
Y.~D. Lensky, J.~Song, P.~Samutpraphoot, and L.~Levitov,
\newblock arXiv:1412.1808 .

\bibitem{sui_2015}
M.~Sui {\em et~al.},
\newblock arXiv:1501.04685.

\bibitem{shimazaki_2015}
Y.~Shimazaki {\em et~al.},
\newblock arXiv:1501.04776.

\bibitem{rahzavy_2003}
M.~Rahzavy,
\newblock {\em Quantum Theory of Tunneling} (World Scientific, New Jersey,
  2003).

\bibitem{peres_2010}
N.~Peres,
\newblock \rmp\ {\bf 82}, 2673 (2010).

\bibitem{dassarma_2011}
S.~{Das Sarma}, S.~Adam, E.~H. Hwang, and E.~Rossi,
\newblock \rmp\ {\bf 83}, 407 (2011).

\bibitem{oka_2009}
T.~Oka and H.~Aoki,
\newblock \prb\ {\bf 79}, 081406(R) (2009).

\bibitem{kundu_2014}
A.~Kundu, H.~A. Fertig, and B.~Seradjeh,
\newblock \prl\ {\bf 113}, 236803 (2014).

\bibitem{gu_2011}
Z.~Gu, H.~A. Fertig, D.~P. Arovas, and A.~Auerbach,
\newblock \prl\ {\bf 107}, 216601 (2011).

\bibitem{HunSanYou13a}
B.~Hunt, J.~D. Sanchez-Yamagishi, A.~F. Young, M.~Yankowitz, B.~J. LeRoy, K.~Watanabe, T.~Taniguchi, P.~Moon, M.~Koshino, P.~Jarillo-Herrero, and R. C. Ashoori,
\newblock Science {\bf 340}, 1427 (2013).

\bibitem{CheShiYan14a}
Zhi-Guo Chen, Zhiwen Shi, Wei Yang, Xiaobo Lu, You Lai, Hugen Yan, Feng Wang, Guangyu Zhang, and Zhiqiang Li,
\newblock Nature Comm. {\bf 5}, 4461 (2014).

\bibitem{mak_2010}
K.~F. Mak, C.~Lee, J.~Hone, J.~Shan, and T.~F. Heinz,
\newblock \prl\ {\bf 105}, 136805 (2010).

\bibitem{mak_2012}
K.~F. Mak, K.~He, J.~Shan, and T.~F. Heinz,
\newblock Nature Nanotech. {\bf 7}, 494 (2012).

\bibitem{arrachea2005}
L.~Arrachea,
\newblock \prb\ {\bf 72}, 125349 (2005).

\bibitem{Sancho}
M.~P. Lop{\'e}z Sancho, J.~M. Lop{\'e}z Sancho, J.~M.~L. Sancho, and J.~Rubio,
\newblock J. Phys. F {\bf 15}, 851 (1985).

%\bibitem{Perry1994}
%M.~D. Perry and G.~Mourou,
%\newblock Science {\bf 264}, 917 (1994).
%
%\bibitem{Backus1998}
%S.~Backus, C.~G. Durfee, M.~M. Murnane, and H.~C. Kapteyn,
%\newblock Rev. Sci. Instrum. {\bf 69}, 1207 (1998).
%
%\bibitem{Cooke2015}
%David Cooke, private communication.

\bibitem{Note1}
These parameters are well within the capabilities of available lasers; David Cooke, private communication. %~\cite{Perry1994,Backus1998,Cooke2015}.

\bibitem{Du2008}
X.~Du, I.~Skachko, A.~Barker, and E.~Y. Andrei,
\newblock Nat. Nanotechnol. {\bf 3}, 491 (2008).

\bibitem{Bolotin2008}
K.~I. Bolotin, K.~J. Sikes, J.~Hone, H.~L. Stormer, and P.~Kim,
\newblock \prl\ {\bf 101}, 096802 (2008).

\bibitem{Xue2011}
J.~Xue, J.~Sanchez-Yamagishi, D.~Bulmash, P.~Jacquod, A.~Deshpande, K.~Watanabe, T.~Taniguchi, P.~Jarillo-Herrero, and B.~J. LeRoy,
\newblock Nat. Mater. {\bf 10}, 282 (2011).

\bibitem{Efetov2015}
D.~K. Efetov %\emph{et al.},
L.~Wang, C.~Handschin, K.~B. Efetov, J.~Shuang, R.~Cava, T.~Taniguchi, K.~Watanabe, J.~Hone, C.~R. Dean, and P.~Kim,
\newblock arXiv:1505.04812.

\bibitem{kohler2005}
S.~Kohler, J.~Lehmann, and P.~H{\"a}nggi,
\newblock Phys. Rep. {\bf 406}, 379 (2005).

\end{thebibliography}

\end{document}